\pdfoutput=1 
\documentclass{llncs}

\usepackage{graphicx}
\usepackage{ifthen}
\usepackage{sidecap}
\usepackage{subfigure}
\usepackage{amsmath}
\usepackage{amssymb}
\usepackage{color}
\usepackage{algorithm}
\usepackage[latin1]{inputenc}
\usepackage[noend]{algpseudocode}

\DeclareMathOperator{\RR}{\bbbr}
\DeclareMathOperator{\NN}{\bbbn}

\newcommand{\Graph}{\ensuremath{\mathcal{G}}}

\begin{document}
\mainmatter  

\title{Maximum Gain Round Trips with Cost Constraints}

\authorrunning{
}   
\author{Franz Graf, Hans-Peter Kriegel, Matthias Schubert}
\institute{
Institute for Informatics, Ludwig-Maximilians-Universit\"at M\"unchen, Oettingenstr.\ 67, D-80538 Munich, Germany\\
\{graf,kriegel,schubert,\}@dbs.ifi.lmu.de\\
}

\maketitle
\begin{abstract}
Searching for optimal ways in a network is an important task in
multiple application areas such as social networks, co-citation
graphs or road networks. In the majority of applications, each
edge in a network is associated with a certain cost and an optimal
way minimizes the cost while fulfilling a certain property, e.g
connecting a start and a destination node. In this paper, we want
to extend pure cost networks to so-called cost-gain networks. In
this type of network, each edge is additionally associated with a
certain gain. Thus, a way having a certain cost additionally
provides a certain gain. In the following, we will discuss the
problem of finding ways providing maximal gain while costing less
than a certain budget. An application for this type of
problem is the round trip problem of a traveler: Given a certain
amount of time, which is the best round trip traversing the most
scenic landscape or visiting the most important sights? In the
following, we distinguish two cases of the problem. The first does
not control any redundant edges and the second allows a more
sophisticated handling of edges occurring more than once. To
answer the maximum round trip queries on a given graph data set,
we propose unidirectional and bidirectional search algorithms.
Both types of algorithms are tested for the use case named above
on real world spatial networks.
\end{abstract}


\section{Introduction}\label{sec:introduction}
Searching for optimal ways in networks is an important task in
many application areas. In most cases, an optimal way is a way
minimizing the cost while fulfilling a certain property. The cost is
usually connected to traversing the edges being part of the way.
For example, in road networks the cost of each edge might be
considered as the time it takes to traverse the edge. Finding the
fastest route between two nodes can now be defined as finding the
route with the property to connect both points and additionally
having a minimum cost. In social network analysis, the cost is
often measured in hops and the shortest path between two members
in the network is considered as a measure for the strength of
their relationship. Other application areas for networks might be
co-citation networks and protein interaction networks where
shortest paths can be used to express similarity. In all of these
domains it is expected that traversing a certain edge is connected
to a certain cost.

In this paper, we like to extend this view by additionally
connecting the traversal of an edge to a certain amount of gain.
In our running example, we consider a traveler who wants to take a
hike in a nature resort. Usually, the traveler wants to see as
much of the landscape as possible. Thus, walking on a scenic trail
with a good view and passing by important landmarks represents
some gain. However, our traveler can usually only walk for a
certain period of time until (s)he wants to be back to the
starting point, e.g. the parking lot. Thus, the task is to
maximize the gain of our traveler while not spending more than a
certain amount of time. Another application for the proposed query
type is a car announcing a blood donation event at the local
hospital. The hospital has hired the car with the speaker for a
limited period of time only and now wants to find a route reaching
as many people willing to donate blood as possible. Let us note
that the problem can be easily extended to searching a way ending
at a different destination than the starting point. Though we will
mainly focus on the case of round trips, the proposed query
processing can be easily extended to the more general case.

A round trip in our definition is a way and thus, it is allowed to
pass by the nodes and edges more than once. Thus, it is different
from finding Euler or Hamilton paths in a graph. An important
issue about our problem definition is the possibility to traverse
the same edge for an arbitrary amount of times. Depending on the
given task, this might lead to round trips having a large
redundancy in the number of visited edges. Therefore, we
distinguish our problem further into two sub problems: The first
allows edges to be visited several times and thus simply maximizes
the gain while keeping the cost below the given threshold $\tau$.
In the second setting, the definition of round trips is extended
by allowing that each edge in the round trip is visited at most
$k$ times. Furthermore, each edge is considered only once when
calculating the gain without consideration of any additional
traversal.

To solve both problems, we will introduce algorithms that
determine a maximum gain round trip for all cost budgets being
smaller than a certain threshold $\tau$. To tackle this
computationally complex task for practically relevant cost
thresholds, we will introduce pruning mechanisms and 
bidirectional search methods. The proposed algorithms are
evaluated by searching round trips on real world map data,
obtained from OpenStreetMap.

The rest of the paper is organized as follows. Section
\ref{sec:related} reviews related work. In section
\ref{sec:formalization}, we define preliminaries and our new
queries. Section \ref{sec:skyline} and section \ref{sec:redundancy}
introduce pruning methods and search algorithms. Section
\ref{sec:experiments} evaluates our algorithms on real world data
w.r.t. runtime and various parameter settings. Finally, section
\ref{sec:conclusion} summarizes the paper and outlines directions
for future research.

\section{Related Work}\label{sec:related}
Common route search which starts from a single source node to at
least one target node is also known as the single-source problem.
This problem has been studied very extensively for a long time
\cite{CheGolRad94,Dan62,DenFox79,Dij59,FreTar87,GalPal88,Gol01,JacMarNag62,Mey01,Tho99,ZhaNoo98}.
Also the task of finding not just the shortest or fastest route
but the top $k$ routes has achieved quite some interest and has
been studied for several years e.g. in \cite{Yen71}.

The closest scenario to the cost-gain networks discussed in this
paper are multi-attribute or multi-cost networks that associate
multiple types of cost to traversing a single edge, e.g. length
and average speed. In \cite{MouLinYiu10}, the authors introduce
preference queries in such multi-cost networks. In particular, the
paper proposes ranking and skyline queries to compute and sort a
result set of possible target destinations in a multi-cost
transport network. This work is different from the work presented
in this paper because the query result consists of possible
destination locations. Another related work is presented in
\cite{SchRenKri10}. In this work, the authors also work on a
multi-cost or multi-attribute network and compute a skyline
operator on the paths leading from a given starting point to a
destination. The query result consists of all paths from the
starting node to the destination node having a pareto optimal cost
vector. The important difference to the queries in this paper
consists in the use of gain attributes. Using gain has a major
impact on the characteristics of the solutions. In cost and
multi-cost networks optimal solutions always consist of shortest
paths w.r.t. some possibly combined cost function. In a cost-gain
network, the impact of traversing an edge needs to be measured
w.r.t. the provided gain as well. Thus, optimal solutions  do not
need to be paths but can be ways. For example, the skyline
operator proposed in \cite{SchRenKri10} cannot generate
interesting round trips because leaving an edge cannot decrease
any cost value compared to the trivial solution of simply staying
a the starting node. Thus, the trivial solution would always
dominate any other way leading away from the starting node.

To the best of our knowledge, there exists no other work that
formulates the search for optimal round trips while employing a
cost-gain network. In theoretical computer science, there exist
methods for finding all cycles present in a graph that will
contain the most interesting round trips or parts of it. For
example, \cite{MehMic09} deals with the task of finding a complete
cycle base in a graph. However, finding a complete cycle base is a
different task and does not consider cost and gain values for the
different edges.

\section{Cost-Gain Networks, Round Trips and Queries}\label{sec:formalization}

A network is represented by a graph where the edges have two
attributes, cost and gain. Thus, we call this graph a cost-gain
network:
\begin{definition}[Cost-Gain Network (CGN)] \\
A cost-gain network is a graph $\Graph(V,E,cost,gain)$ where $V$
is denoting the set of vertices and $E \subset V\times V$ is
denoting the set of edges. \\
$cost: E \rightarrow \RR^0_+$ is called
cost function where $cost(e)$ denotes the non-negative cost for an
edge $e \in E$. \\
$gain: E \rightarrow \RR^0_+$ is called gain
function where $gain(e)$ denotes non-negative gain for an edge $e \in
E$.
\end{definition}

In our running example, a CGN represents a network of roads,
streets, paths, sidewalks and trails. The nodes correspond to
crossings, the edges correspond to path segments in the graph. The
cost of a segment represents a certain combination of the
characteristics of this segment, like the length, the maximum
speed, the time needed to pass the segment, the inclination etc.
The gain of a segment can be defined correspondingly by combining
all characteristics that the user considers as beneficial, e.g. a trail
that is not used by cars might be considered as more attractive
for a hiker than a highway segment etc.

Let us also note that the attributes of an edge $e_i=(n_s,n_d)$
need not be the same as edge $\hat{e}=(n_d,n_s)$ even though both
edges describe the exactly same path just in different directions
as the user might for example avoid declines of a certain degree
whilst accepting inclines of some other degree. Also the impact on
travel speed for this edge might be different.

\begin{definition}[way] A way $w$ is a sequence of edges
$((v_{1},v_{2})$, $(v_{2},v_{3})$,$\ldots$, $(v_{k-1},v_{k}))$
where the following condition holds:
\begin{equation}\label{eq:path1}
\forall 1 \leq i < k : \exists e \in E: e = \left( v_{i}, v_{i+1} \right)
\end{equation}
The cost of a way $w$ is defined as follows:
\begin{equation}
cost(w) = \sum_{i=1}^{k-1}{cost((v_i, v_{i+1}))}
\end{equation}
The gain of a way $w$ is defined as follows:
\begin{equation}
gain(w) = \sum_{i=1}^{k-1}{gain((v_i, v_{i+1}))}
\end{equation}

\end{definition}

In other words, a way is a sequence of connected edges. A round
trip is a way starting and ending with the same node $s$.
\begin{definition}[round trip] A way \\
$w=((v_{1},v_{2}),( v_{2},v_{3}),\ldots, (v_{k-1},v_{k}))$ is
called round trip if $v_1 = v_k$.
\end{definition}

In the above definition of a round trip, there is no consideration
of whether or not a round trip contains one and the same edge more than
once. However, passing the same edge very often does not yield an
additional benefit in most applications. For example, for a
hiker looking for the most scenic walk, it might be necessary to
pass the same edge twice, if some edges having a large gain are
placed in a dead end. However, after visiting a spot and returning
on the same path, it would not make any sense to go there again.
Thus, a possibility to control the redundancy is to limit the
number of times an edge can be contained in the round trip.
Furthermore, traversing the same edge more than once should not
contribute to the gain of the round trip. Thus, we can extend our
definition of round trips to round trips with redundancy control.
Formally, this can be formulated as follows:

\begin{definition}[Redundancy Control] Given the \\
round trip $r$ in the CGN $G(V,E,cost,gain)$, $r$ is called a
round trip under redundancy control with level $k \in \NN$, if
the following condition holds:
\begin{displaymath}
\forall (v_{a}, v_{b}) \in r: \{ (v_i,v_{i+1})\in r|(v_i = v_a
\wedge v_{i+1} = v_b) \vee (v_i = v_b \wedge v_{i+1} = v_a ) \} |
\leq k
\end{displaymath}
For a round trip $r$ under redundancy control, the gain is
calculated as follows:
\begin{displaymath}
gain(r) = \sum_{e \in ES(r)}{gain((v_i, v_{i+1}))} \ \ \mbox{with}
\ \ ES(r) = \{ e \in r \}
\end{displaymath}
\end{definition}

After defining both types of round trips, we can now define
maximum gain round trip queries:

\begin{definition}[Maximum Gain Round Trip Query] \\
Given the CGN $G(V,E,cost,gain)$, a starting node $s \in V$ and
cost threshold $\tau \in \RR^+$, the result of a maximum gain
round trip query (MGRQ) is the set $R$ of round trips, such that
for each element $r \in R$ the following constraints hold: (a)
$cost(r) \leq \tau$\\
(b) $\forall r,\hat{r} \in R :
gain(r)<gain(\hat{r})\Leftrightarrow cost(r) < cost(\hat{r})$
\end{definition}

In other words, the result of a MGRQ contains a round trip
providing maximum gain for each cost level being smaller than
$\tau$. Thus, the result set contains all pareto optimal ways
starting and ending at $s$.

\section{Algorithms for MGRQs}\label{sec:skyline}
In this section, we will examine the problem of searching maximum
gain round trips with cost constraints without controlling the
redundancy.

\subsection{Pruning Round Trips}

The simplest pruning condition for a way $w=((v_1,_2)$, $\hdots$,
$(v_{k-1}$,$v_k))$ during our search is that the way cannot be
extended into a round trip $r=((v_1,_2), \hdots,
(v_{k-1},v_k),\hdots$, $(v_l,v_1))$ with $cost(r)\leq \tau$. Thus,
if $cost(w) > \tau$, $w$ can be pruned because the cost of any
extension of $w$ must be larger than $cost(w)$.

\begin{figure}[t]
  \centering
  \subfigure[ ]{
    \label{fig:pathEx1}
    \includegraphics[width=0.45\textwidth]{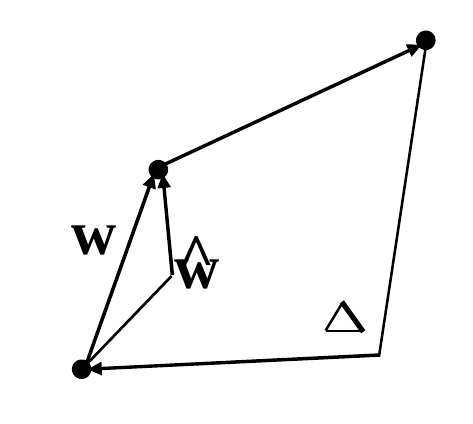}
  }
  \subfigure[ ]{
    \label{fig:pathEx2}
    \includegraphics[width=0.45\textwidth]{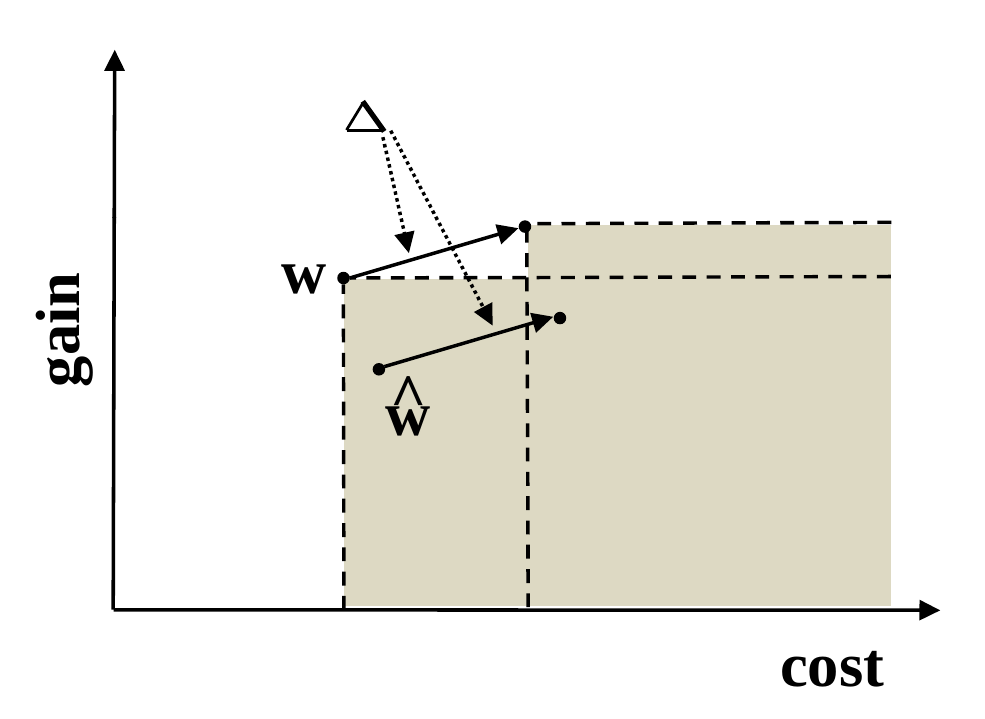}
  }
  \caption{Local pruning of pareto optimal paths: if $\hat{w}$ is dominated by $w$,
  any further extension $\Delta$ of $w$ and $\hat{w}$ does not change the pruning
   relation so that $\hat{w}+\Delta$ is still dominated by $w+\Delta$.}
  \label{fig:pathExtension}
\end{figure}

Let us note that pruning w.r.t. gain is not that simple.
Extending a way usually increases the gain and thus, there is no
upper limit for the gain of a way. However, since we are only
looking for round trips having an optimal cost-gain ratio, we can
use the following observation for defining a further pruning
criterion: 
For each $r \in R$, we can guarantee that there is no other round
trip generating more gain and having at most the same cost.
Therefore, each partition $w$ of the result round trip $r$ has a
pareto optimal cost gain ratio w.r.t all other ways starting and
ending at the same position as $w$. The intuition behind this
conclusion is that each part of the round trip serves three
purposes: it moves the traveler to the end of the way, it
generates a certain amount of cost and it provides a certain
amount of gain. Apart from this, the properties of the way do not
influence the properties of the complete round trip. Formally, we
can define the following pruning rule based on the domination
relationship:

\begin{lemma} [local pruning by domination]
Let $w = ((v_1,v_2),\hdots,(v_{k-1},v_k))$ be a way in
$G(E,V,cost,gain)$, then $w$ is dominated by another way
$\hat{w}=((v_1,\hat{v_2})$,$\hdots$, $(\hat{v_{k-1}}$, $v_k))$ if
either $cost(w) > cost(\hat{w}) \wedge gain(w) \leq gain(\hat{w})$
or $cost(w) \geq cost(\hat{w}) \wedge gain(w) < gain(\hat{w})$. If
$w$ is dominated by $\hat{w}$, then $w$ cannot be extended into an
element of the result set of an MGRQ for $v_1$.
\begin{proof}
Consider the result round trip $r$ with $gain(r)$ and $cost(r)$.
Due to the definition of the result set, we can rule out that
there is another round trip $\hat{r}$ having at most the same cost
and more gain or at least the same gain and less cost. Now,
consider a way $w=(v_1,v_2,\hdots,v_i)$ being part of $r$. If
there exists another way $\hat{w}=(v_1,\hat{v_2}, \hdots, v_i)$ in
$r$ leading from $v_1$ to $v_i$ with $cost(\hat{w})<cost(w) \wedge
gain(\hat{w})\geq~gain(w)$ or $cost(\hat{w}) \leq cost(w) \wedge
gain(\hat{w})\geq~gain(w)$, then it is possible to construct a
round trip $r_{new}$ by replacing $w$ by $\hat{w}$ in $r$.
However, in this case $\hat{r}$ would dominate $r$ which
contradicts the condition that $r$ is part of the result set.
\end{proof}
\end{lemma}

An illustration of this pruning mechanism can be found in figure
\ref{fig:pathExtension}.

\subsection{A Basic MGEQ Algorithm}
In our descriptions, we will denote the edges starting at node $v$
as outlinks of $v$ while the edges ending at $v$ are called
inlinks. Our algorithm employs two data structures. The first is a
hash table called \textit{node tab} containing an entry for each
visited node $v_i$. Furthermore, the node tab stores all
undominated ways starting at $s$ and ending at $v_i$. For each of
these ways, we store a flag indicating whether we already
processed the way in a previous step or not. In the following, we
will use the expression ''update the node tab with way $w$'' for
the following steps being used in our algorithm:
\begin{enumerate}
\item Check whether $w$ is dominated by any entry of the node tab.

\item If $w$ is not dominated, insert $w$ into the node tab entry.

\item Remove all entries from the node tab which are dominated by
$w$.

\end{enumerate}
Our second data structure is a priority queue containing all
nodes. Each node $v_i$ is prioritized by the maximum gain among
all ways ending at $v_i$ and the queue is organized in descending
order.

\begin{figure}[h!]
\centering \fbox{
  \begin{minipage}{0.96\textwidth}
  \sf\scriptsize
   SimpleMGRQ(Node $s$, Float $\tau$)\\
   (1)\quad\hspace*{0em}Nodetab  tab  = InitNotetab()\\
   (2)\quad\hspace*{0em}PriorityQueue queue = InitQueue()\\
   (3)\quad\hspace*{0em}FOR EACH Link $l$ IN $s.outlinks()$ DO \\
   (4)\quad\hspace*{1em}Way $w$ = new $Way(s,l)$
   (5)\quad\hspace*{1em}$tab.update(w)$\\
   (6)\quad\hspace*{1em}$queue.update(w.last,w.gain)$\\
   (7)\quad\hspace*{0em}END FOR\\
   (8)\quad\hspace*{0em}WHILE NOT $queue.isEmpty()$ DO\\
   (9)\quad\hspace*{1em}Entry $entry =queue.pop()$\\
   (10)\quad\hspace*{1em}List$<$Way$>$ aktList = entry.getUndominated()\\
   (11)\quad\hspace*{1em}aklist = removeProcessedWays(aktlist) \\
   (12)\quad\hspace*{1em}setProcessed(aktlist)\\
   (13)\quad\hspace*{1em}List$<$Way$>$ candidates =
   extendWays(aktList)\\
   (14)\quad\hspace*{1em}FOR EACH $w$ IN candidates DO \\
   (15)\quad\hspace*{2em}IF $w.cost < \tau$ DO \\
   (16)\quad\hspace*{3em}$tab.update(w)$\\
   (17)\quad\hspace*{3em}$queue.update(w.last,w.gain)$\\
   (18)\quad\hspace*{2em}END IF\\
   (19)\quad\hspace*{1em}END FOR\\
   (20)\quad\hspace*{0em}END WHILE\\
   (21)\quad\hspace*{0em}RETURN $tab.getEntry(s)$\\
\end{minipage}
} \caption{Pseudocode of the simple MGRQ Algorithm}
\label{code:algorithm1}
\end{figure}

The algorithm starts by generating the ways resulting from
following all out links of the starting node $s$. Afterwards the
ways are used to update the node tab as well as the queue. Now the
algorithm enters the main loop which is repeated until the
priority queue is empty. In each iteration, the algorithm pops the
top node from the queue and retrieves all unprocessed ways from
the node tab. Let us note that we have to keep already processed
ways in the node tab for determining locally dominated ways.
However, it is note required to process each way more than once.
Now, each unprocessed way is marked as processed. Afterwards we
extend each way by all of its out links generating a set of
candidate ways. The candidate ways are checked whether their cost
exceeds the limit $\tau$. Afterwards each candidate
$c=((s,v_1),\hdots,(v_k,v_{new}))$ is checked against the ways
being stored in the node tab entry of their end node $v_{new}$. If
$c$ is not dominated by any other way, it is inserted into the
node tab entry. Furthermore, if $c$ dominates former members of
the node tab entry, these members can be pruned due to their sub
optimal cost gain ratio. If the maximum gain of any node tab entry
being modified is increased, the entry has to be updated in the
queue. After the queue is empty, the result of our query can be
found in the node tab entry of the starting node $s$. Figure
\ref{code:algorithm1} displays the algorithm in pseudo code.

Let us note that the above algorithm is capable to find arbitrary
pareto optimal ways having a cost less than $\tau$ and ending at
any visited node. Thus, it is not restricted to the search of
round trips.

\begin{figure}[h!]
\centering \fbox{
  \begin{minipage}{0.96\textwidth}
  \sf\scriptsize
     BidirectionalMGRQ(Node $s$, Float $\tau$)\\
   (1)\quad\hspace*{0em}Nodetab  tab  = InitNotetab();\\
   (2)\quad\hspace*{0em}PriorityQueue queue = InitQueue()\\
   (3)\quad\hspace*{0em}FOR EACH Link $l$ IN $s.outlinks()$ DO \\
   (4)\quad\hspace*{1em}Way $w$ = new $Way(s,l)$\\
   (5)\quad\hspace*{1em}$tab.updateStart(w)$\\
   (6)\quad\hspace*{1em}$queue.update(w.last,w.gain)$\\
   (7)\quad\hspace*{0em}END FOR\\
   (8)\quad\hspace*{0em}FOR EACH Link $l$ IN $s.inlinks()$ DO \\
   (9)\quad\hspace*{1em}Way $w$ = new $Way(s,l)$\\
   (10)\quad\hspace*{1em}$w=w.reverse()$\\
   (11)\quad\hspace*{1em}$tab.updateReturn(w)$\\
   (12)\quad\hspace*{1em}$queue.update(w.first,w.gain)$\\
   (13)\quad\hspace*{0em}END FOR\\
   (14)\quad\hspace*{0em}WHILE NOT $queue.isEmpty()$ DO\\
   (15)\quad\hspace*{1em}Entry $entry =queue.pop()$\\
   (16)\quad\hspace*{1em}List$<$Way$>$ fwdList = entry.getundominatedStart()\\
   (17)\quad\hspace*{1em}fwdList = removeProcessed(fwdList)\\
   (18)\quad\hspace*{1em}setProcessed(fwdList)\\
   (19)\quad\hspace*{1em}List$<$Way$>$ bwdList = entry.getundominatedReturn()\\
   (20)\quad\hspace*{1em}bwdList = removeProcessed(bwdList)\\
   (21)\quad\hspace*{1em}setProcessed(bwdList)\\
   (22)\quad\hspace*{1em}FOR EACH $w$ IN fwdList  DO \\
   (23)\quad\hspace*{2em}IF $w.cost > \frac{\tau}{2}$ DO\\
   (24)\quad\hspace*{3em}$fwdList.delete(w)$\\
   (25)\quad\hspace*{2em}END IF\\
   (26)\quad\hspace*{1em}END FOR \\
   (27)\quad\hspace*{1em}List$<Ways>$ fwdCandidates =
   extendFwdWays(fwdList)\\
(28)\quad\hspace*{1em}FOR EACH $w$ IN fwdCandidates DO\\
(29)\quad\hspace*{2em}$tab.updateStart(w)$\\
(30)\quad\hspace*{2em}$queue.update(w.last(),w.gain)$\\
(31)\quad\hspace*{1em}END FOR \\
(32)\quad\hspace*{1em}List$<$Ways$>$ bwdCandidates =
extendbwdWays(bwdList)\\
(33)\quad\hspace*{1em}FOR EACH w IN bwdCandidates DO \\
(34)\quad\hspace*{2em}IF $w.cost < \frac{\tau}{2}$ DO\\
(35)\quad\hspace*{3em}$tab.updateReturn(w)$\\
(36)\quad\hspace*{3em}$queue.update(w.last,w.gain)$\\
(37)\quad\hspace*{2em}END IF\\
(38)\quad\hspace*{1em}END FOR\\
(39)\quad\hspace*{0em}END WHILE\\
(40)\quad\hspace*{0em}LIST$<WAYS>$ result = new
List<ways>()\\
(42)\quad\hspace*{0em}FOR EACH entry IN tab DO\\
(43)\quad\hspace*{1em}FOR EACH startWay IN
entry.getundominatedStart() DO\\
(44)\quad\hspace*{2em}FOR EACH retWay IN
entry.getundominatedReturn()  DO\\
(45)\quad\hspace*{3em}Way roundtrip = startWay.extend(retWay)\\
(46)\quad\hspace*{3em}$result.update(roundtrip)$ \\
(47)\quad\hspace*{2em}END DO\\
(48)\quad\hspace*{1em}END DO\\
(49)\quad\hspace*{0em}END FOR\\
(50)\quad\hspace*{0em}Return result.entries
\end{minipage}
} \caption{Pseudocode of the bidirectional MGRQ Algorithm}
\label{code:bidirectional}
\end{figure}

\subsection{Bidirectional Round Trip search}
A further method to improve the runtime is bidirectional search as
employed in the well-known bidirectional Dijkstra search for
shortest paths \cite{SinDeC77}. For searching maximum gain round
trips, bidirectional search yields an even stronger advantage. The
algorithm described above generates ways having a cost of at most
$\tau$. Thus, it has to visit any node that is reachable by
spending the cost limit $\tau$. However, a round trip has to end
at its starting node. Obviously, examining a way $w$ having a
network distance of $\tau$ cannot have a return path $w_2$ to the
starting node that would not exceed $\tau$. Thus, it is only
necessary to explore ways $w_1$ for which there exists a return
path $w_2$ having at most a cost of $cost(w_2)=\tau-cost(w_1)$.

To conclude, it is only necessary to extend each way $w$ until
$cost(w)$ exceeds $\frac{\tau}{2}$. In particular, we can
distinguish two cases, when trying to split $w$ into two
partitions of equal cost. In the first case, there is a node
after the distance of exactly $\frac{\tau}{2}$. In the second
case, the split point having exactly the cost of $\frac{\tau}{2}$
is located at the edge $e=(v_i,v_{i+1})$. Then, there is a unique
partitioning of $w$ into three parts $w_1$,$e$ and $w_2$ and by
extending $w_1$ by $e$, we will get a unique partitioning of $w$
into starting way $w_1$ and return way $w_2$. Based on this
observation, we can stop extending ways that exceed the cost limit
of $\frac{\tau}{2}$ at most by one hop and thus, approximately
work with only half of the search radius.

Though bidirectional search is applicable for searching any
maximum gain way $w$ having $cost(w)<\tau$, it is especially well
suited for searching round trips. Since the area of the graph that
has to be explored for finding the starting ways is the same as
the area being explored for finding the return ways, it is
possible to simultaneously search for both parts of a round trip.
Thus, the part of the graph being accessed during query processing
is significantly decreased.

Our bidirectional search algorithm employs a node tab storing
pareto optimal ways leading to a visited node $v$ managing two
lists of undominated ways. The first contains all undominated ways
$w_i$ leading from the starting node $s$ to $v$ and the second
manages all undominated ways starting at $v$ and leading to $s$.
Updating the node tab is used in an almost identical way as
described above. The only difference to the above use is that we
have to distinguish whether $w$ is a starting way or a return way
and update the corresponding list. The processing order is again
managed by a priority queue that is ordered by the maximum gain
being observed in either part of the node tab.

The algorithm proceeds as follows: At initialization, the
algorithm considers all outlinks of the starting node $s$ and
generates a first set of starting ways. After updating the node
tab with these ways and inserting the corresponding nodes into the
priority queue, we use all inlinks to $s$ to generate a first set
of return ways and again update the node tab and the priority
queue. Now the algorithm enters the main loop which is iterated
until the priority queue is empty. In each iteration $i$, the
algorithm pops the top node $v_i$ from the queue. Afterwards, it
retrieves all unprocessed ways from the list of starting ways
leading to $v_i$ in the node tab and checks if these have a cost
smaller than $\frac{\tau}{2}$. If a way $w$ is still smaller than
$\frac{\tau}{2}$, the algorithm extends $w$ by all outlinks of
$v_i$ and generates a candidate set $C_w$. Each element of $c \in
C_w$ is now used to update the node tab and the priority queue in
case the maximum gain of the node tab entry of the last node $c$
is increased. Afterwards the algorithm retrieves all unprocessed
return ways. Each of these ways $w_{ret}$ is extended by all
inlinks to a set of candidate ways $C_{w_{ret}}$ of ways starting
at $v_i$ and ending at the origin $s$. Then each candidate $c \in
C_{w_{ret}}$ is checked whether its cost is still less or equal
$\frac{\tau}{2}$. If $c$ passes this test, $c$ is used to update
the node tab and the queue, if the maximum gain of the entry of
its first node is increased. Let us note that it is important to
check starting ways and return ways at different stages of the
processing to achieve all ways sufficing the partitioning
described above. After the priority queue is empty, i.e. there
is no unprocessed way left that can be extended any further, we
need to join the pareto optimal starting ways and return ways. The
result set is organized in a list of undominated round trips which
is updated by visiting all entries of the node tab. For each entry
representing node $v$, we examine all combinations of undominated ways
starting at $w_{start}$ and return ways $w_{ret}$. For each pair of
ways $w_{start}$ and $w_{ret}$, we first of all determine the cost
of the corresponding round trip by adding
$cost(w_{start})+cost(w_{ret})$ and the corresponding gain by
adding $gain(w_{start})+gain(w_{ret})$. Based on this cost-gain
vector, we can now update the result list of undominated round
trips. If the new round trip is undominated, we join both ways and
add the result to the result list. If the new round trip even
dominates formerly pareto optimal round trips, the now dominated
round trips are deleted. After each entry of the node tab is
processed, the algorithm terminates and the result consists of all
pareto optimal round trips with a cost of up to $\tau$. Figure
\ref{code:bidirectional} describes the bidirectional search in
pseudo code.

\begin{figure}[t]
  \centering
  \subfigure[ ]{
    \label{fig:localPruning1}
    \includegraphics[width=0.16\textwidth]{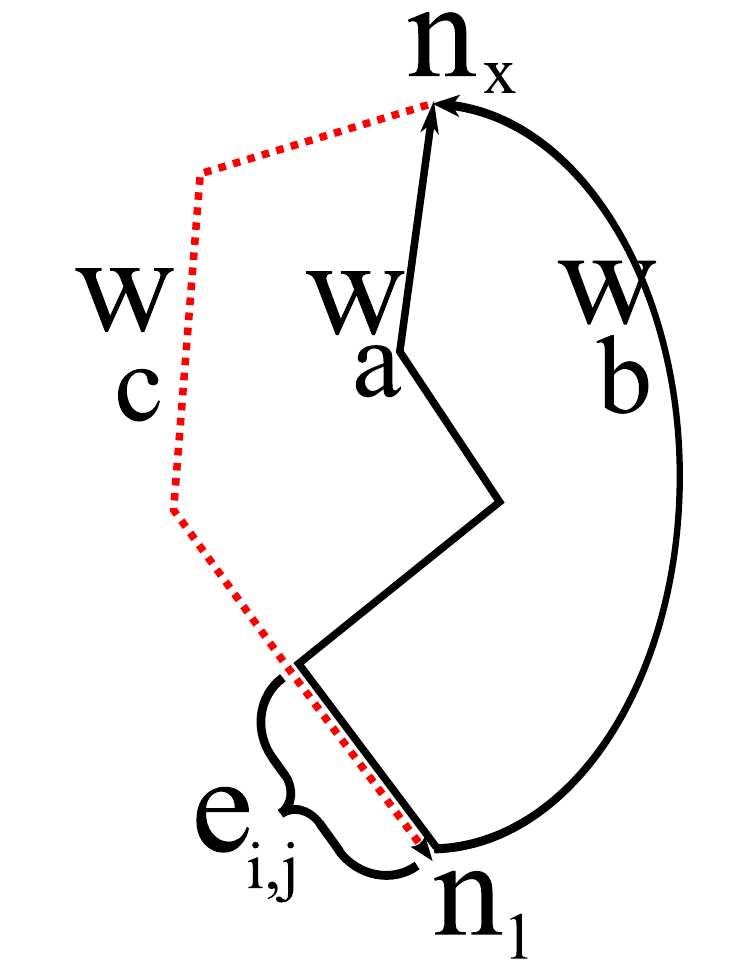}
  }
  \subfigure[ ]{
    \label{fig:localPruning2}
    \includegraphics[width=0.16\textwidth]{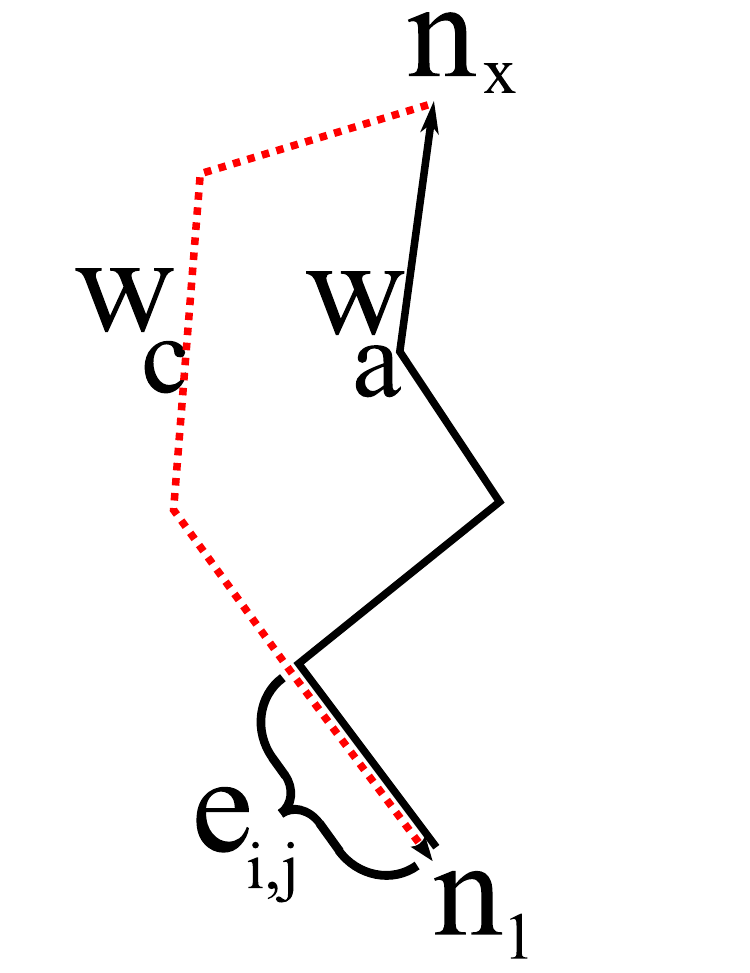}
  }
  \subfigure[ ]{
    \label{fig:localPruning3}
    \includegraphics[width=0.16\textwidth]{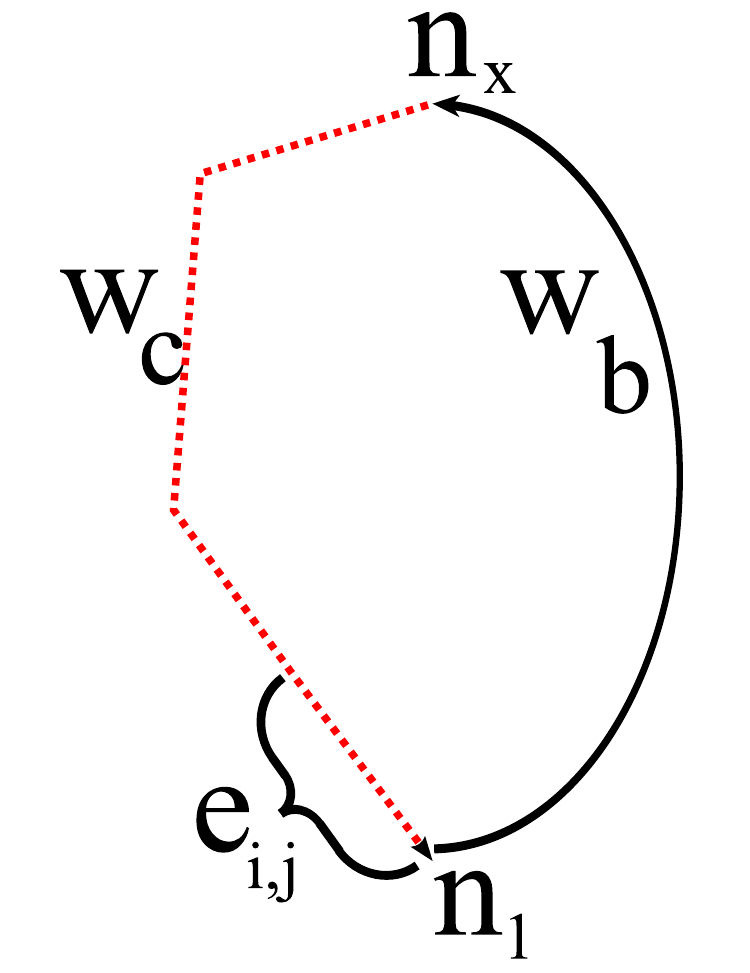}
  }
  \subfigure[ ]{
    \label{fig:localPruning4}
    \includegraphics[width=0.33\textwidth]{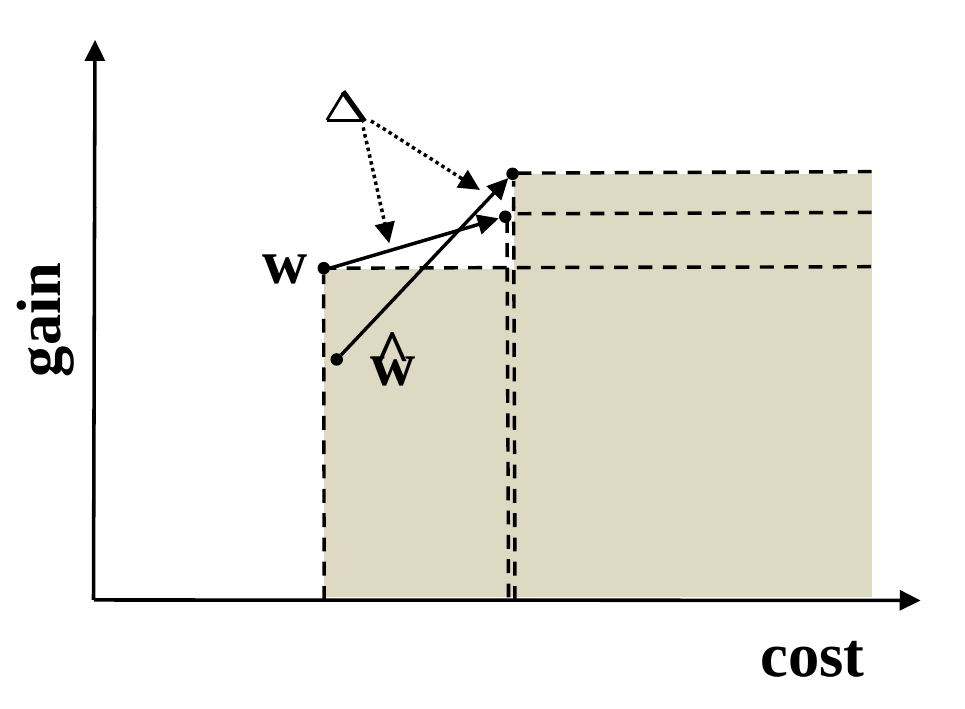}
  }
  \caption{Figure that indicates the problem of local pruning if an edge cardinality is constrained. $\Delta$ indicates the cost/gain of $p_c$ which is added to the paths $p_a$,$p_b$. The cost contribution to $p_a$ and $p_b$ is the same with different gain because $p_c$ and $p_a$ share an edge. And thus, its pruning power is lowered as well.}
  \label{fig:localPruning}
\end{figure}

\section{MGRQs with Redundancy Control}
\label{sec:redundancy} In this section, we will discuss maximum
gain round trip queries under the constraints limiting the amount
and the impact of edges which occur more than once. In particular,
we will not allow that a round trip contains the same edge more
than $k$ times. Furthermore, as the cost will sum up over
duplicate edges as well, the gain of ways is calculated over the
set of all edges being contained in the round trip. Since there
are no duplicates in a set, each edge can add its gain only once
to the round trip. Please recall that we will consider edge $(v_i,v_j)$ as equivalent to $(v_j,v_i)$ w.r.t. redundancy
control.

\subsection{Pruning and Redundancy Control}
Since computing the cost is the same in both types of round trips,
pruning ways w.r.t.  their cost is applicable in the same way as
described in Section \ref{sec:redundancy}. However, when trying to
exploit the cost-gain ratio to prune ways, both redundancy control
mechanisms have a major impact. When using redundancy control, we
cannot guarantee that all partitions of a pareto optimal round
trip $r$ are pareto optimal in the same ways as described above.
Figure \ref{fig:localPruning} illustrates this effect in a simple
example. Even though the way $w$ dominates the way $\hat{w}$, $w$
is not part of the pareto optimal round trip $r$. However, the
dominated way $\hat{w}$ is part of $r$ and replacing $\hat{w}$ by
$w$  would lead to a round trip $\hat{r}$ being dominated by $r$.
In this example we can observe that the influence of $w$ to the
gain of the complete round trip $r$ is not limited to the
$cost(w)$, $gain(w)$ and its end node $v$. If $w$ is extended into
a round trip, all edges being visited on $w$ will not add any gain
and thus, these edges influence the cost gain ratio of the
complete round trip. A similar observation holds for limiting the
cardinality of each edge in a round trip to $k$. In this case, a
pareto optimal round trip $r$ might contain the dominated way $w$
because the way $\hat{w}$ dominating $w$ contains edges that would
occur more than $k$ times when replacing $w$ by $\hat{w}$ in $r$.
In other words, the corresponding round trip $\hat{r}$ containing
$\hat{w}$ would violate the redundancy control.

In order to retain the possibility of local pruning, the
domination relation defined above
must be extended by the set of visited edges. Thus, to extend
local pruning to redundancy controlled round trips, we can
formulate the following lemma:
\begin{lemma}[dominaton under redundancy control] \\
Let $\hat{w}=((v_1,\hat{v}_2),\hdots,(\hat{v}_{k-1},v_k)$ be a way
in $G(E,V,cost,gain)$, then $\hat{w}$ is called dominated under
redundancy control by another node $w =
((v_1,v_2),\hdots,(v_{k-1},v_k))$  if the following conditions
hold:\\
(a)$ (cost(\hat{w}) > cost(w) \wedge gain(\hat{w}) \leq gain(w))$\\
\hspace*{0.5cm} $\vee (cost(\hat{w}) \geq cost(w) \wedge gain(\hat{w}) <gain(w)$ \\
(b)For $ES(w) = \{ e \in E | e \in w \}$,$ES(\hat{w}) =
\{ e \in E | e \in \hat{w} \}$: $ES(w) \subseteq ES(\hat{w})$\\

If $\hat{w}$ is dominated under redundancy control by $w$, then
$\hat{w}$ cannot be extended into an element of the result set of
an MGRQ for $v_1$.
\begin{proof}
The proof for condition (a) is identical to the proof in section
\ref{sec:skyline}. It remains to show that replacing $\hat{w}$ in
a round trip $\hat{r}$ by $w$ cannot lead to a reduced gain due to
duplicate edges or an invalid round trip. Consider the way
$w_{ret}$ extending $\hat{w}$ to the way $\hat{r}$. The additional
gain being earned by traversing $w_{ret}$ in $\hat{r}$ can be
described as
\begin{displaymath}
gain(w_{ret})-\sum_{e \in ES(\hat{w})\cap ES(w_{ret})}{gain(e)}
\end{displaymath}
Since $gain(e) \geq 0$ and $ES(w) \cap ES(w_{ret}) \subseteq
ES(\hat{w}) \cap ES(w_{ret})$, $gain(r) \geq gain(\hat{r})$.
Furthermore, if $\hat{r}$ does not violate the redundancy
parameter $k$, then $r$ cannot violate it either because it
follows from $ES(w) \subseteq ES(\hat{w})$ that $w$ does not
contain any edge $e$ with $e \not \in \hat{w}$. Thus, any
violation in $r$ would be encountered in $\hat{r}$ as well.
\end{proof}
\end{lemma}

A major issue for the usefulness of this lemma is whether there
are enough ways that can be pruned to justify the additional
effort for comparing the edge sets. In the following, we start off
by discussing the worst case scenario and afterwards point out the
cases in which our pruning rule still justifies the overhead.

\begin{figure}[t]
    \centering
        \includegraphics[width=0.4\textwidth]{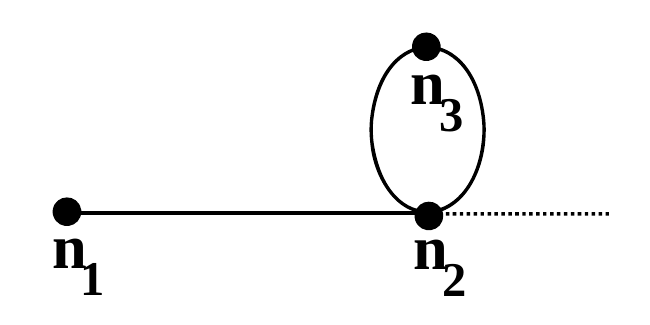}
    \caption{This figure depicts a way $(n_1,n_2,n_3,n_2)$ that can be pruned if $gain((n_2,n_3,n_2))=0$.
    In any case, additional traversals of the cycle $(n_2,n_3,n_2))$ do not yield any gain. Thus,
    the way $(n_1,n_2,n_3,n_2,n_3,n_2)$ would be pruned in any case.}
    \label{fig:prunedways}
\end{figure}

For values of $k \leq 2$ and  $gain(e) > 0$ for each $e \in E$, it
is impossible that any way can be pruned. Since the dominated way
$\hat{w}$ must contain at least the same edges as $w$,
$gain(\hat{w}) \geq gain(w)$.  Due to $w$ dominating $\hat{w}$, we
know $gain(w)=gain(\hat{w})$ and $cost(w) < cost(\hat{w})$. Based
on this observation, it follows that $w$ and $\hat{w}$ must have
the same set of edges, i.e. $ES(w)=ES(\hat{w})$ because $\forall e
\in E: gain(e) >0$ and $gain(w)=gain(\hat{w})$. Thus, $\hat{w}$
cannot contain an additional edge to $w$. Thus, the only allowed
difference between $\hat{w}$ and $w$ is that $\hat{w}$ visits 
some edges $e \in ES(w)$ more than once. For $k=2$ the only
possibility that $w$ and $\hat{w}$ end with the same node is $w =
\hat{w}$.

Correspondingly, for $k>2$ and $\forall e  \in E: gain(e) \geq 0$,
there might be ways $\hat{w}$ being dominated by $w$. In figure
\ref{fig:prunedways}, we illustrate the types of ways being
pruned. On the left side $\hat{w}$ extends $w$ by a cycle
consisting of new edges offering no gain. The example on the right
hand side excludes the way $\hat{w}$ because it visits a cycle in
$w$ more than once. To subsume, domination under redundancy
control prunes all ways containing cycles which do not provide
gain.

A final pruning mechanism we have to consider is the redundancy
parameter $k$. Since we can prune all ways violating the
cardinality threshold $k$, the number of ways we have to consider
for further extension is usually much smaller than in the general
case without redundancy control. The smaller the value of $k$ is
chosen the stronger is its pruning power.

To conclude, searching for maximum gain round trips under
redundancy control can employ cost-based pruning as in Section
\ref{sec:skyline}. Additionally, choosing a small parameter value
$k$ also has the power to prune invalid ways during the traversal.
However, when increasing the value of $k$, the number of pruned
paths is strongly decreasing. For parameter settings,
domination-based pruning justifies the overhead because it
prevents the algorithms from exploring ways which are revisiting
identical parts of the way multiple times.

\subsection{Algorithm for MGRQs with Redundancy Control}
Both algorithms for answering MGRQs described in section
\ref{sec:skyline} are easily adaptable for the case of redundancy
controlled round trips. To modify the algorithms, we first of all
have to integrate the redundancy control. Therefore, the gain
calculation has to be changed in order to prevent duplicate edges to
add any further gain. Furthermore, each time the algorithm tries
to extend a way $w$ by an edge $(v_i,v_j)$, we have to check
whether $(v_i,v_j)$ or its complement $(v_j,v_i)$ is already
contained in $w$ more than $k$ times. This way, all invalid
candidates are already pruned before they are constructed in the
first place.

In our case, we want to use domination under redundancy control as
described above. The method for checking domination must be
extended by additionally checking for the subset condition.
However, if we employ small values of $k$ and the majority of
edges has a non-zero gain, it often makes sense to abandon pruning
based on domination to avoid the computational overhead. In this case, the list
of ways in the node tab entry for node $v$ contains all valid ways
found so far.

A final difference for the first algorithm described in Section
\ref{sec:skyline} is that the result set has to be stored in a
dedicated list of pareto optimal ways. In the previous setting,
the node tab entry of the starting node $s$ already contains a
pareto optimal list of round trips. However, since the pruning
rules under redundancy control are less restrictive, the remaining
dominated round trips must be removed before returning the result
set. Let us note that this difference has no impact to the
bidirectional algorithm because the last step joining starting
ways and return ways has to construct a new pareto optimal result
set anyway. After implementing the named modifications, we can
employ both algorithms to compute MGRQs under redundancy control.

\section{Experimental Evaluation}\label{sec:experiments}
In our evaluation, we use data obtained from OpenStreetMap\footnote{Data and Map data \copyright ~OpenStreetMap (and)
contributors, CC-BY-SA http://www.OpenStreetMap.org}. We
preprocessed the data using the converter provided in
\cite{GraKriRenSch10} to remove some of the nodes having a degree
of 2. In our tests, we examined three different areas which are
popular for hiking: Kirchsee (GER),  Jasper(AL,CA) and Grand
Canyon Village(AZ,US). For each area, we selected a central 
starting point\footnote{Kirchsee:
http://www.openstreetmap.org/?node=312519650\\ Jasper:
http://www.openstreetmap.org/?node=915165849\\ Grand Canyon:
http://www.openstreetmap.org/?node=174618876}.


For all our tests, we chose the Euclidean distance
between two nodes as a cost criterion. To represent the gain, we
considered the road type. Thus, we assigned each edge allowing a
maximum speed of less than 30\,km/h with the gain of 1 and for the
rest of the edges, we assigned a gain of 0.

\paragraph{Processing time.}
In our first set of experiments, we compare the runtime of the
round trip search with redundancy control (\emph{rtSearchwRC}) to
the search allowing redundancy (\emph{rtSearch}). Furthermore,
we compare the bidirectional search (\emph{bidirectionalRTS}) and
the bidirectional search with redundancy control (\emph{bidirectionalRTSwRC}). For \emph{rtSearchwRC} and
\emph{bidirectionalRTSwRC} the redundancy parameter $k$ was set to
1.

\begin{figure}[h!]
  \centering
  \subfigure[Kirchsee]{
    \label{fig:timekirchsee}
    \includegraphics[width=0.75\textwidth]{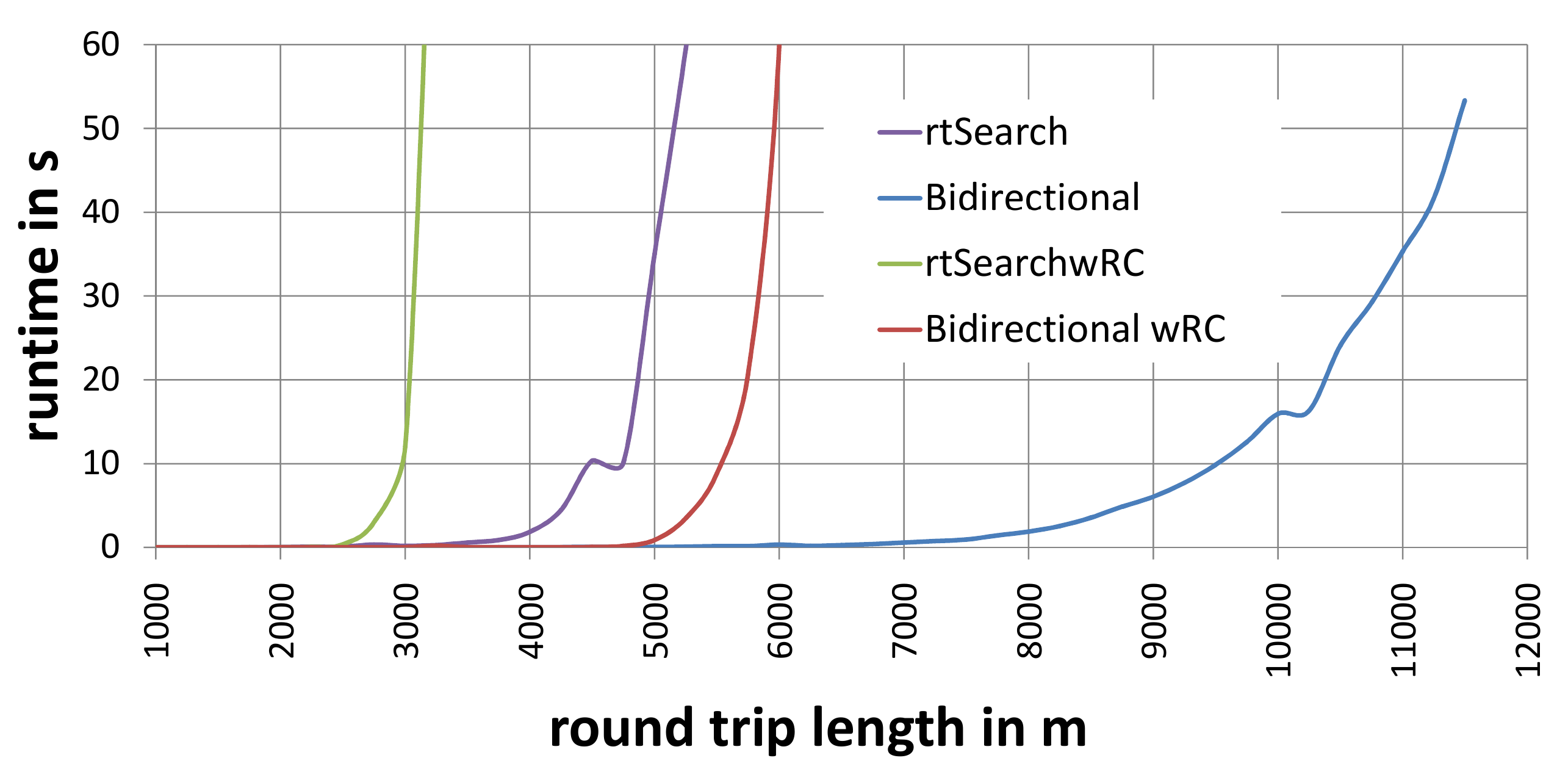}
  }
  \subfigure[Jasper (AL,CA)]{
    \label{fig:timeJasper}
    \includegraphics[width=0.75\textwidth]{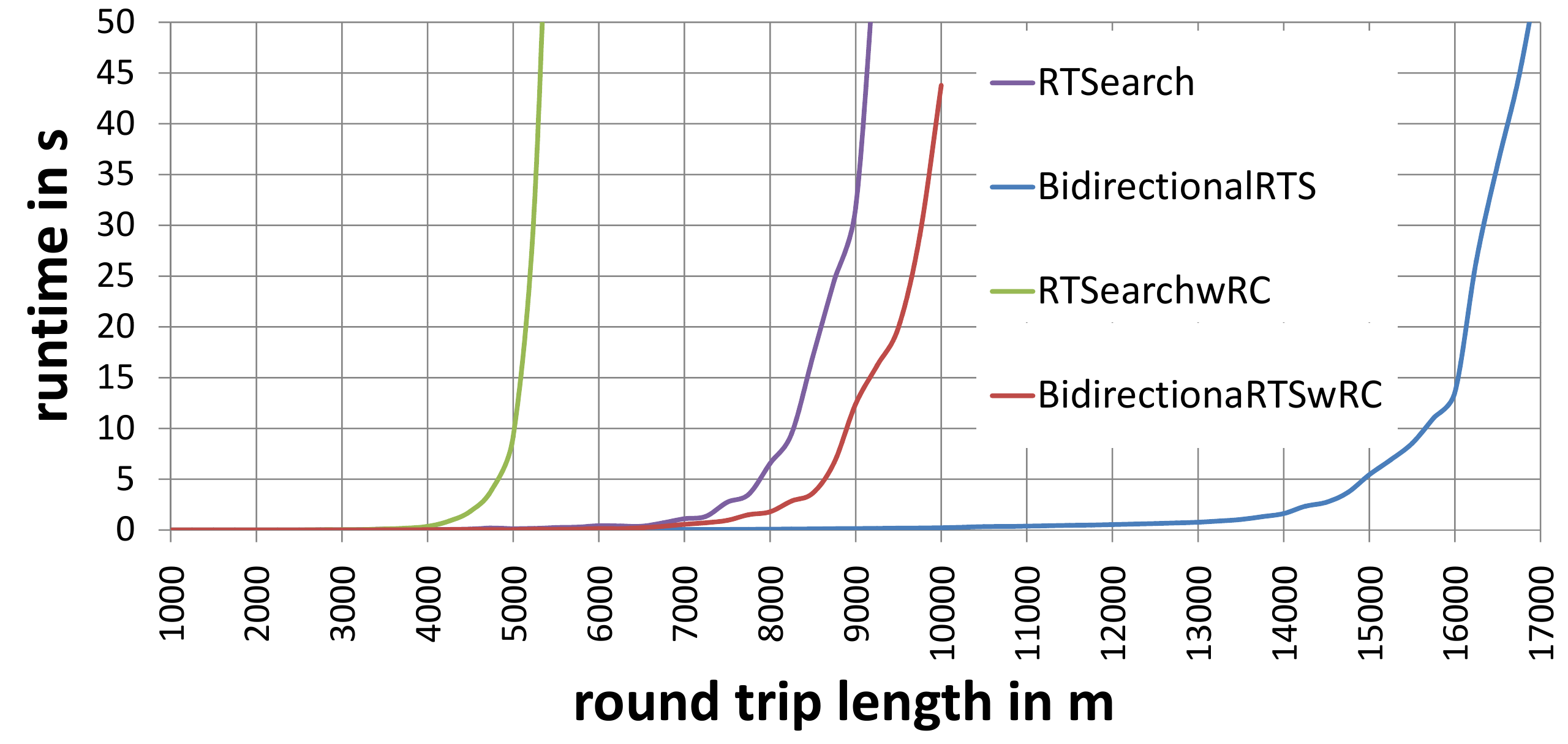}
  }
  \subfigure[Grand Canyon Village (AZ,US)]{
    \label{fig:timeGrandCanyon}
    \includegraphics[width=0.75\textwidth]{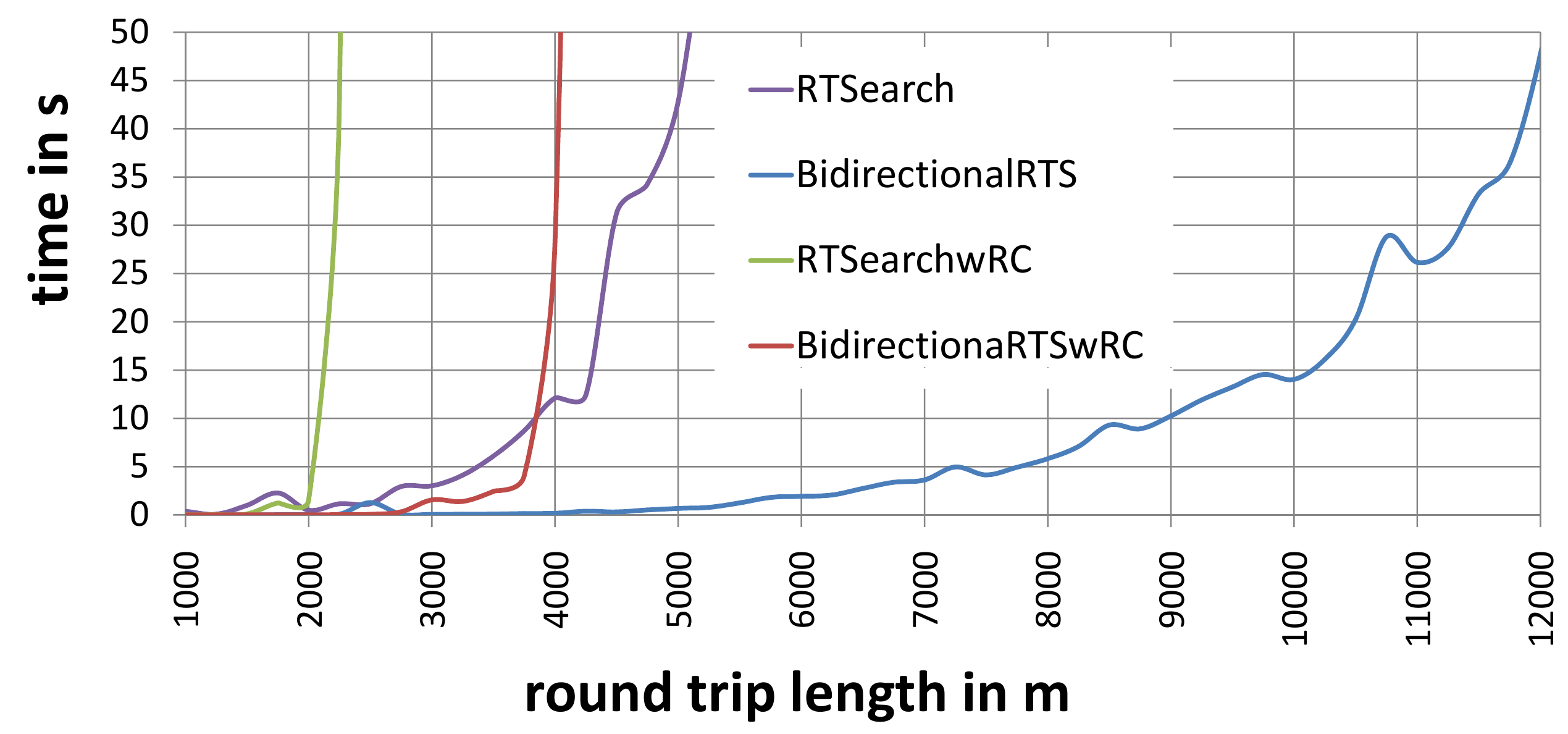}
  }
  \caption{Figures \ref{fig:timekirchsee}, \ref{fig:timeGrandCanyon} and \ref{fig:timeJasper} show the runtime of the proposed algorithms. For \emph{rtSearchwRC} and \emph{bidirectionalRTSwRC}
  $k$ was set to 1.}
  \label{fig:resulttime}
\end{figure}

Figure \ref{fig:resulttime} displays the runtime of all four
algorithms with increasing cost values for all three maps. A first
observation is that all algorithms display a certain cost budget
for which the search starts to show a super linear increase in
search time. However, we can observe that for employing
bidirectional search the threshold for which query processing
takes less than a minute can be extended to a reasonably large
distance for round trips, large enough to be interesting (up
to 17\,km). Furthermore, on all graphs the bidirectional search
could extend the cost limit being processable in less than one minute
in comparison to the unidirectional algorithm for the same query
type. A final conclusion that can be drawn from the results is
that queries without redundancy control can be processed much
faster than those employing redundancy control. This observation
can be explained by the fact that dominance on the cost-gain graph
is a much stronger pruning mechanism than dominance under
redundancy control. Later on, we will present an experiment
showing that dominance under redundancy control is very important
for larger values of $k$.

\begin{figure}[t]
  \centering
  \subfigure[BidirectionalRTS]{
    \label{fig:nodes}
    \includegraphics[width=0.47\textwidth]{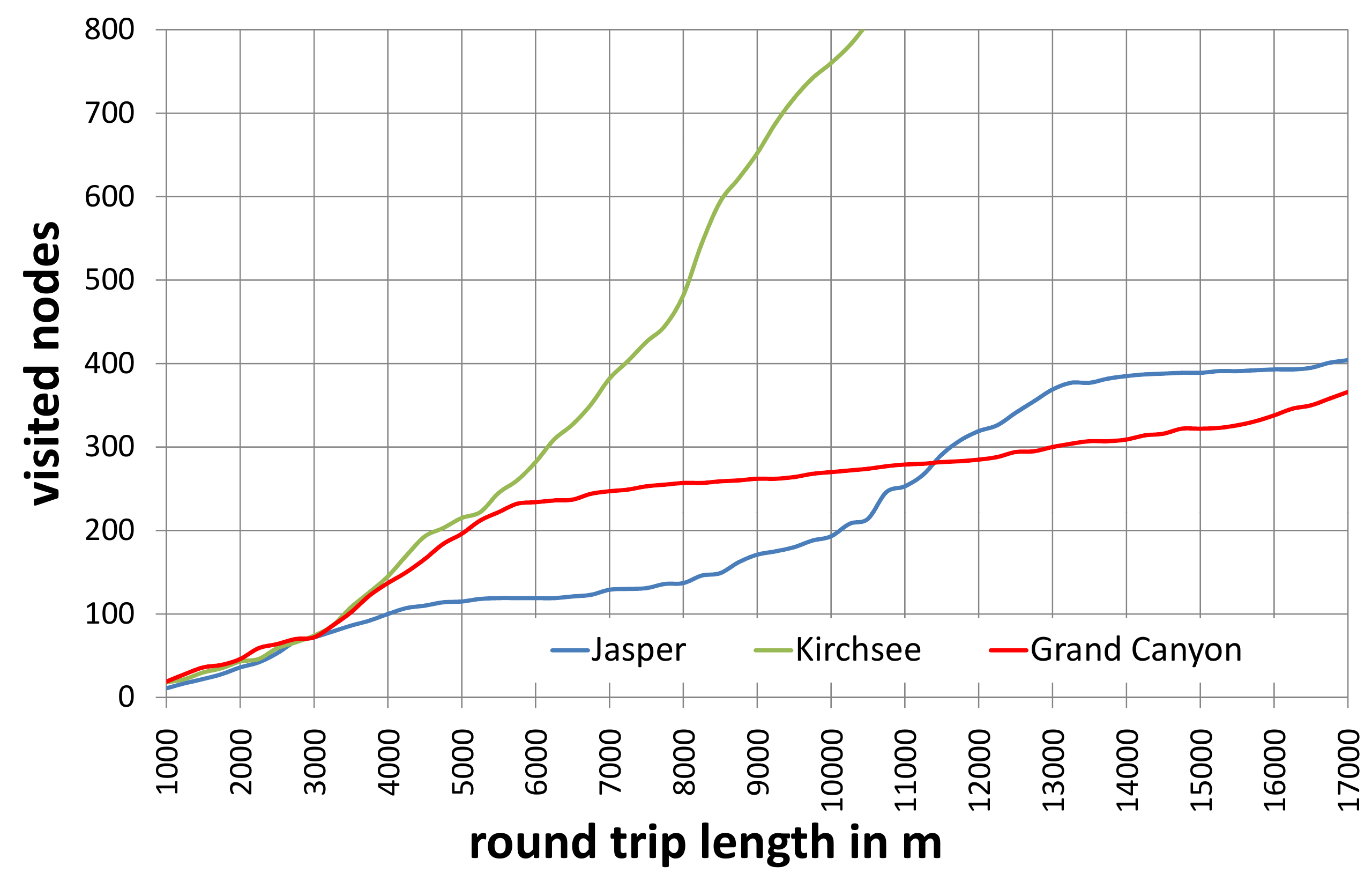}
  }
  \subfigure[BidirectionalRTSwRC]{
    \label{fig:nodeswRC}
    \includegraphics[width=0.47\textwidth]{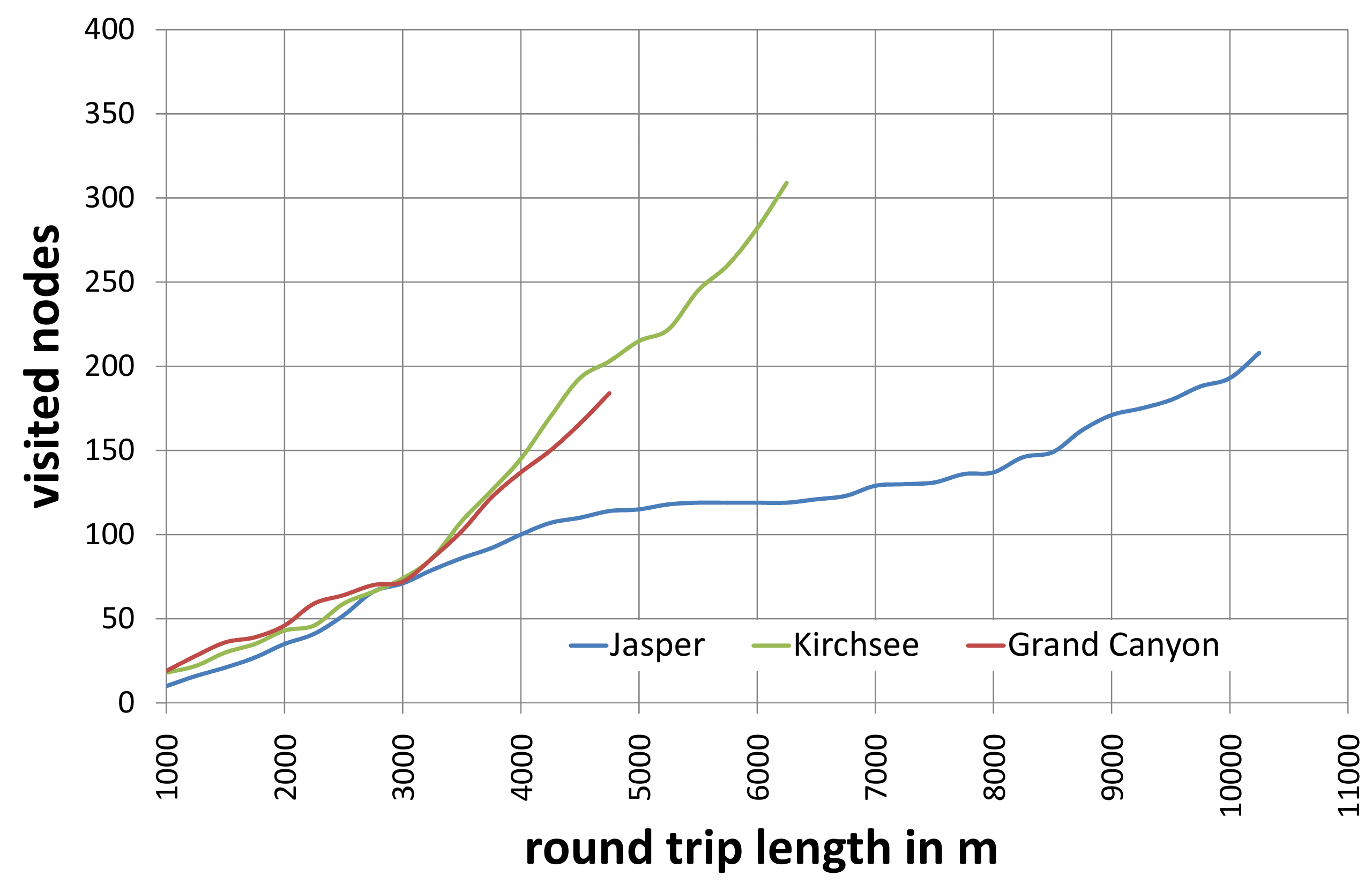}
  }
  \caption{Number of nodes visited by BidirectionalRTS (\ref{fig:nodes}) and BidirectionalRTSwRC
  (\ref{fig:nodeswRC}) for all three maps.}
  \label{fig:VisitedNodes}
\end{figure}

\paragraph{Search space.}
Another important factor when searching ingraph data is the portion
of the graph which has to be available in main memory. Thus we
examined the increase of the nodes visited during the search
with \emph{bidirectionalRTS} and \emph{bidirectionalRTSwRC} w.r.t.
the cost threshold $\tau$. The results for all three maps and
both bidirectional search algortihms is displayed in figure
\ref{fig:VisitedNodes}. It can be seen that the portion of the
graph visited increases approximately linearly with the
threshold parameter $\tau$. Furthermore, we can observe that
\emph{bidirectionalRTS} and \emph{bidirectionalRTSwRC} visit 
comparable portion of the graph. Let us note that
\ref{fig:nodeswRC} is scaled differently. Since the 
maximum cost threshold processable for
\emph{bidirectionalRTSwRC} is smaller than for
\emph{bidirectionalRTS}, we could not generate values for all
distance thresholds we displayed in figure \ref{fig:nodes}. To
conclude, MGRTQs visit rather small portions of the graph. The high
complexity of MGRTQs is rather caused by the large amount of
possible round trips than by the size of the graph.

\begin{figure}
    \centering
        \includegraphics[width=0.75\textwidth]{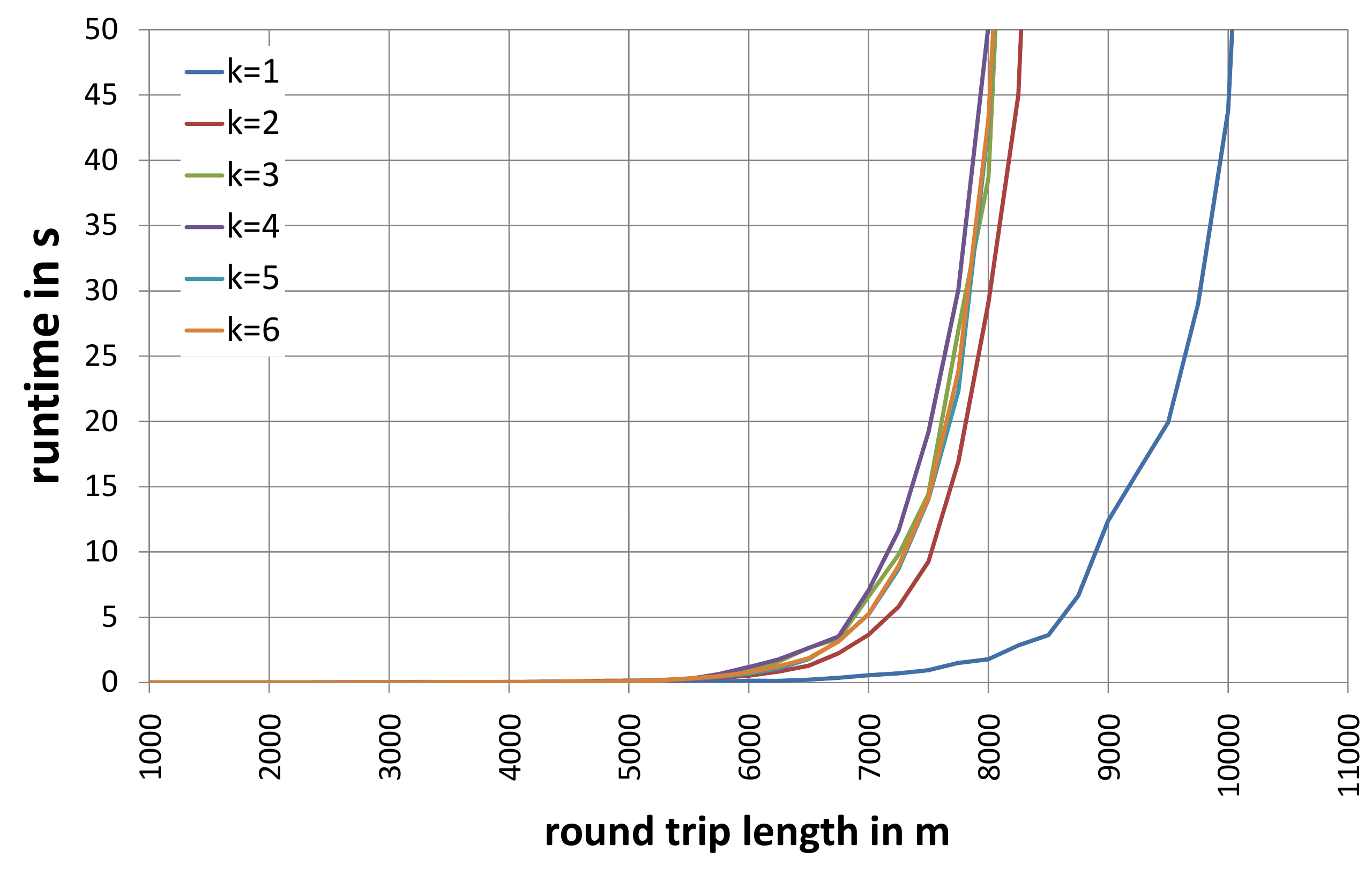}
    \caption{Runtime of \emph{bidirectionalRTSwRC} for different values of $k$ }
    \label{fig:K1-6}
\end{figure}

\paragraph{Impact of the Redundancy Level $k$.}
Another impact factor on the search speed when using redundancy
control is $k$, which limits the cardinality of an edge. Figure
\ref{fig:K1-6} illustrates the impact of $k$ on the retrieval
times of \emph{bidirectionalRTSwRC} on the Jasper map with varying
 $k$ from 1 to 6. At first it can be observed that setting
 $k$=1 displays better runtimes and thus, the distance threshold
 which can still be computed ended at 10\,km. For values of $k\
\geq 2$, the maximum threshold which could be reached was around 8\,km. An interesting result is that for all $k \geq 2$ the run time
was approximately the same. Thus, with the exception of the special case
$k=1$ the value of $k$ does not have a strong influence on
the runtime. The reason for this effect can be found in the
pruning rule employing dominance under redundancy control. We will
demonstrate this effect more clearly in the next experiment.

\begin{figure}
    \centering
        \includegraphics[width=0.75\textwidth]{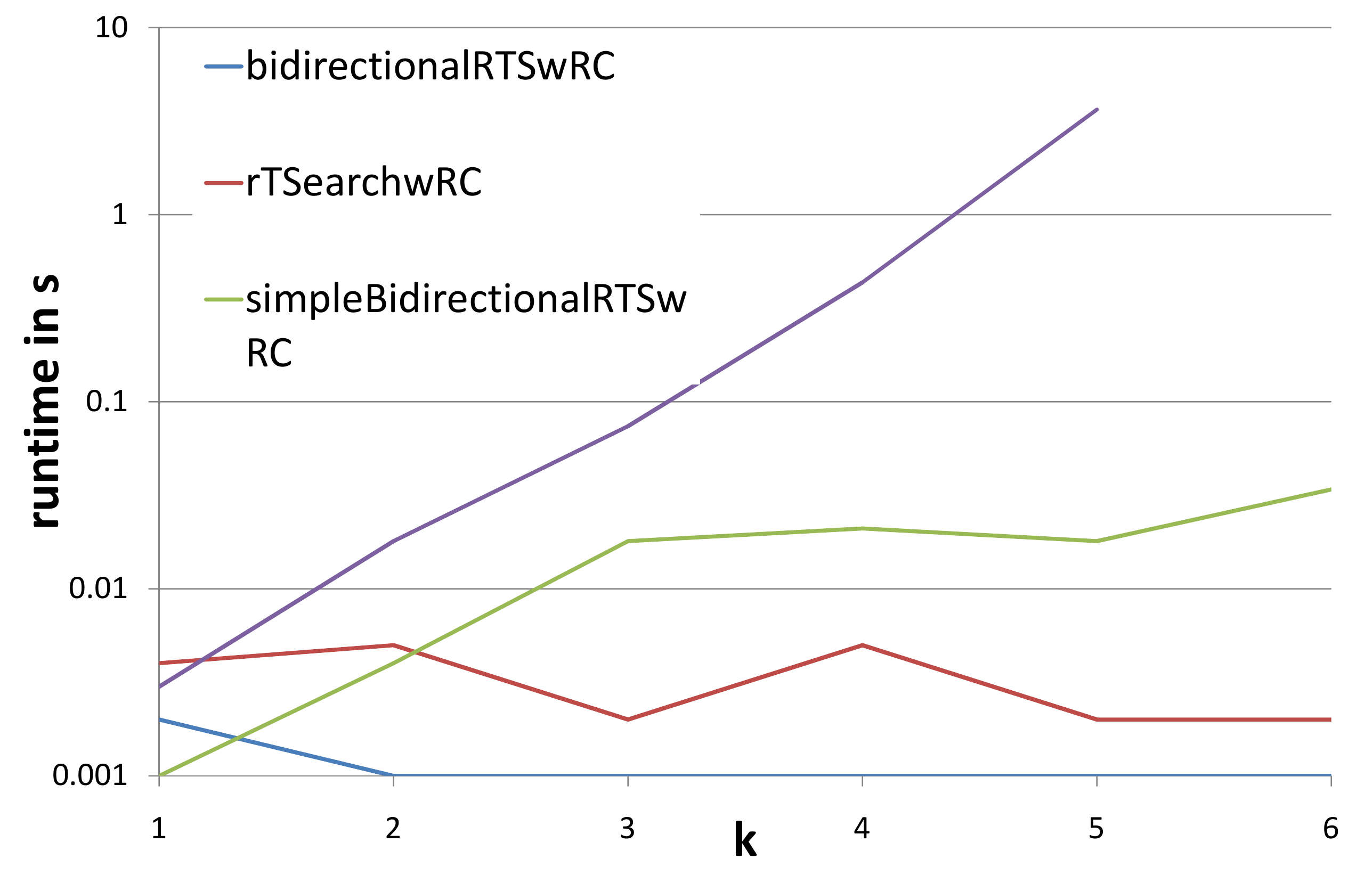}
    \caption{Impact of redundancy control on runtime when executed on the Jasper map with $\tau = 1\,700\,m$ and $k$ varying from 1 to 6.  }
    \label{fig:redundancyControl}
\end{figure}


\paragraph{Domination under Redundancy Control.}
In Section \ref{sec:redundancy}, we proposed domination under
redundancy control to implement an additional pruning rule. We
already explained why the impact of this rule strongly depends on
the value of the redundancy parameter $k$. In the following, we
will examine the runtime behaviour of our search algorithms with
dominance pruning (\emph{rtSearchwRC}) and
(\emph{bidirectionalRTSwRC}) and search algorithms without the
dominance pruning rule \emph{simpleRTSwRC} and
\emph{simpleBidirectionalRTSwRC} for increasing values of $k$. The
result is displayed in figure \ref{fig:redundancyControl}. The
experiment was executed on the Jasper map with $\tau = 1\,700\,m$
and $k$ varying from 1 to 6. Let us note that the relatively small
value for $\tau$ was chosen to still be able to compute results
with \emph{simpleRTSwRC} and \emph{simpleBidirectionalRTSwRC} for
larger values of $k$.

In case of $k = 1$, our pruning rule cannot have any impact. Thus,
it can be seen that the algorithms without the additional pruning
rule perform better for values of $k\ =1$. For increasing values
of $k$, we can observe an exponential increase of the runtime
for the basic algorithms \emph{simpleRTSwRC} and
\emph{simpleBidirectionalRTSwRC}. In contrast, \emph{rtSearchwRC}
and \emph{bidirectionalRTSwRC} show an almost constant runtime
behaviour for increasing values of $k$. Thus, employing dominance
pruning under redundancy control is a significant improvement for
larger values of $k$.

\section{Conclusions}\label{sec:conclusion}
In this paper, we examined maximum gain round trip queries (MGRQs)
with cost constraints in cost-gain networks. A cost-gain network
is a graph where each edge is connected to a certain amount of
cost and additionally provides a certain amount of gain. A round
trip is a way starting and ending at the same node and a maximum
gain round trip is a round trip providing the maximum gain for a
certain cost. The result set of an MGRQ is a set of round trips
containing a maximum gain round trip for every cost level being
less or equal to a cost limit $\tau$. We propose solutions for two
sub problems. The first deals with round trips where edges might
occur multiple times. The second sub problem restricts the number of times
an edge can occur in the solution to a maximum value of $k$. Our
algorithms are tested on real world map data taken from Open Street
Map.  For future work, we plan to examine further pruning
mechanisms which are based on optimistic forward approximations
similar to A*-search. Furthermore, we plan to examine parallel
algorithms to extend the cost limit being still computable.
Furthermore, we will develop approximative algorithms for the case
that an exact search requires too much resources.

\section*{Acknowledgments}
\begin{footnotesize}
This research has been supported in part by the {THESEUS} program in the
{CTC} and {MEDICO} projects. It is funded by the German Federal Ministry of
Economics and Technology under the grant number {01MQ07020}. The
responsibility for this publication lies with the authors.
\end{footnotesize}

\bibliographystyle{abbrv}
\bibliography{RoundtripSearch}
\end{document}